\documentstyle[11pt,gh2001-asp,twoside,epsf]{article}
\def\proptosim{\mathrel{\hbox{\rlap{\hbox{\lower4pt\hbox{$\sim$}}}\hbox{$\propto$}}}}
\begin{document}
%
%
\title{Structure, Kinematics, and Dynamics of Bulges}
\author{M.~Bureau} 
\affil{Hubble Fellow, Columbia Astrophysics Laboratory, 550~West
120th Street, 1027~Pupin Hall, Mail Code~5247, New York, NY~10027,
U.S.A.}
%
%
\begin{abstract}
A historical review of our understanding of bulges is first presented,
highlighting similarities and differences between bulges and
ellipticals. Then, some topics of current interest are reviewed,
bypassing stellar population questions and focusing on structural and
dynamical issues relating bulges and disks. The topics are: i) the
fundamental plane; ii) the evidence for two classes of bulges,
$R^{1/4}$ and exponential, and its significance for bulge formation;
iii) the three-dimensional structure of bulges, in particular the
relation between boxy/peanut-shaped bulges and bars; iv) the nuclear
properties of bulges, and their possible effects on bulge dynamics and
secular evolution; and v) the large-scale mass distribution and
evidence for dark matter in bulges. To conclude, new prospects offered
by wide-field integral-field spectroscopy and other instrument
developments (space and ground-based) are discussed.
\end{abstract}
%
%
\vspace*{-3mm}
\section{Introduction}
Both figuratively and literally, bulges are central to our
understanding of most disk galaxies. The so-called ``super-thins''
apart, bulgeless disks with extreme axial ratios (Karachentsev,
Karachentseva, \& Parnovskij 1993), most disks possess a central
spheroidal-like component called the bulge (better defined as the
central excess over the inward extrapolation of the outer exponential
disk; Carollo, Ferguson, \& Wyse 1999). In fact, the bulge-to-disk
ratio is a defining property of the Hubble classification for
spirals. In many cases, the bulge dominates the central potential, it
determines the position of major resonances, and bulges are now known
to be intimately linked to massive central black holes (BHs; Magorrian
et al.\ 1998). Bulges thus play an active role in the structure,
dynamics, and evolution of the galaxies in which they are embedded.

Writing an exhaustive review in the space allocated is impossible, so
only selected topics related to the structure, kinematics, and
dynamics of bulges will be discussed. Issues regarding stellar
populations will mostly be left out, and clues to bulge evolution
shall be noted only when directly relevant to the issue at hand. The
common thread will be to emphasize similarities and differences
between ellipticals and bulges on the one hand, and bulges and disks
on the other. Topics to be discussed include a brief historical review
(\S~\ref{sec:history}), the fundamental plane (FP) of bulges
(\S~\ref{sec:FP}), their light distribution (\S~\ref{sec:light}), 3D
structure (\S~\ref{sec:3D}), nuclear properties
(\S~\ref{sec:nuclear}), and large-scale mass distribution
(\S~\ref{sec:large-scale}). The conclusions (\S~\ref{sec:conclusions})
will illustrate the perspectives offered by panoramic integral-field
spectrographs and new or recent instruments.
%
%
\section{Bulges and Ellipticals: A Historical Point-of-View\label{sec:history}}
\vspace*{-6mm}
It is fair to say that our basic understanding of the structure and
dynamics of bulges and ellipticals stems from work done in the 1970s
and early 1980s. A few references will be used to illustrate key
developments in the field.

Until the late 1970s, ellipticals and bulges were thought to be very
similar because of similar light profiles, stellar content, and
kinematics. Hints of scaling relations were also beginning to
appear. de Vaucouleurs (1958) showed that Andromeda's\index{object,
M31} bulge followed the $R^{1/4}$ surface brightness profile of
ellipticals. Faber (1977) showed that ellipticals and bulges (S0s and
Ss) had similar stellar populations, sharing the same colors,
Mg$_2$--$M_B$, and NaD--Mg$_2$ relations. Faber \& Jackson (FJ; 1976)
discovered a projection of the FP ($L\proptosim\sigma^4$), shared by
both ellipticals and bulges (taken here as S0s and M31), and showing
no discontinuity in mass-to-light ratio $M/L$. Of course, other
properties were known to differ. From their axial ratio distributions,
Sandage, Freeman, \& Stokes (1970) noted that ellipticals are
consistent with a large range of intrinsic flattening, $q\equiv
b/a\approx1.0-0.3$, while lenticulars and spirals must all have
intrinsically flat disks, with a nearly constant axial ratio
($q\approx0.25$). Furthermore, some of the above similarities were
later proven too simplistic.

Because of the flattened light distribution of ellipticals, oblate
spheroidal models with isotropic velocity dispersions were first
constructed (e.g. Lynden-Bell 1962, Prendergast \& Tomer 1970). Wilson
(1975) showed that, based on the Ostriker \& Peebles (1973) criterion,
models flatter than E4 would be unstable to axisymmetric
perturbations. The crucial developments, however, came with the first
rotation curves based absorption lines, able to probe the stellar
kinematics. Illingworth (1977) demonstrated in a distance-independent
way that ellipticals have only $1/3$ (on average) the rotation
required by oblate isotropic models. Rotation thus contributes little
to their total kinetic energy. Schechter \& Gunn (1979) extended these
results and showed that some ellipticals possess significant
minor-axis rotation and isophotal twists. It was thus clear that
ellipticals were triaxial ellipsoids (or oblate spheroids) flattened
by anisotropic velocity dispersions, not rotation. Binney (1976, 1978)
had already shown that anisotropies could be preserved in the collapse
of a (non-spherical) protogalactic cloud.

These results prompted a new look at bulges. It was quickly realized
that many bulges are not well represented by $R^{1/4}$ light profiles,
some even showing an exponential decline (e.g.\ NGC4565\index{object,
NGC4565}; Jensen \& Thuan 1982). Bulges also appeared to be rotating
more rapidly than the bright ellipticals studied so far (Kormendy \&
Illingworth 1982), suggesting formation through dissipational collapse
rather than merging (the spin parameter $\lambda$ is much larger than
expected from tidal torques, 0.3 rather than 0.05). Nevertheless,
Davies et al.\ (1983) showed that low luminosity ellipticals rotate as
fast, both being consistent with rotationally flattened isotropic
oblate models (see Fig.~1). This clearly suggested a continuum in the
structure and dynamics of spheroids as the luminosity is decreases.

Lynden-Bell (1967) provided a physical basis for the aforementioned
models, showing that violent relaxation (i.e.\ rapid collapse) leads
to a distribution function
$f(E,L_z)\propto\exp(-$cst.$E\,+\,$cst.$L_z)$, where $E$ and $L_z$ are
the specific energy and angular momentum around the symmetry axis, and
the $L_z$ term is non-zero for rotating systems only. In fact, for a
certain concentration index, King's (1966) models reproduce well the
$R^{1/4}$ law over a large range of radii.
%
%
\vspace*{-3mm}
\section{Fundamental-Plane of Bulges\label{sec:FP}}
\vspace*{-2mm}
Both ``early'' bulges and ellipticals follow a common relation, more
general than the FJ relation, known as the FP: $\log
R_e=\alpha\log\sigma+\beta\log I_e+\gamma$, where $R_e$ is the
effective radius, $\sigma$ the central velocity dispersion, $I_e$ the
effective surface brightness, and $\alpha$, $\beta$, and $\gamma$
constants (Djorgovski \& Davis 1987, Dressler et al.\ 1987). If the
systems are in virial equilibrium (as expected), have a constant
$M/L$, and form a homologous family (i.e.\ their properties scale
simply with luminosity or mass), we expect $\alpha=2$ and $\beta=-1$
($\gamma$ varies with distance). In practice, the coefficients differ
from the virial expectation and depend somewhat on their definitions
and measurements. J\o rgensen, Franx, \& Kj\ae rgaard (1996; Fig.~2)
obtain $\alpha=1.24$ and $\beta=-0.82$ in Gunn~$r$, with a scatter of
0.07 in $\log R_e$ (17\% error on individual distances). Crucial for
determining distances, the slope is constant among clusters
(independent of richness, $T_{gas}$, $\sigma_{cluster}$, etc). The
scatter is real, higher for S0s (bulges), and is unlikely due to disks
or projection effects (residuals uncorrelated with the shape of the
light distribution).

%
%
\begin{figure}
\plotone{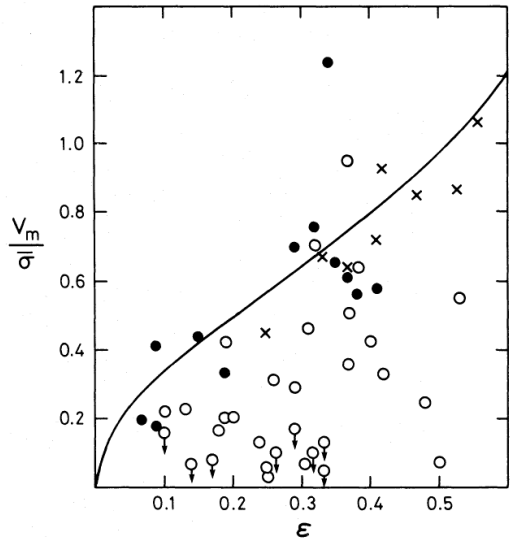}
\caption{Rotational support of spheroids. The ratio of the maximum
velocity of rotation to the average value of the dispersion within
$\onehalf\, R_e$, $V_m/\bar{\sigma}$, is plotted as a function of the
projected ellipticity near $R_e$, $\epsilon$. The solid line
represents isotropic oblate rotator models (Binney 1978). {\em Open
circles:} Bright ellipticals ($M_B\leq-20.5$). {\em Filled circles:}
Faint ellipticals ($M_B>-20.5$). {\em Crosses:} Bulges. Both bulges
and faint ellipticals are consistent with isotropic oblate spheroid
models. Reproduced with permission from Davies et al.\ (1983).}
\end{figure}

Departures from the virial FP can be assigned to a varying
mass-to-light ratio, $M/L\propto
R_e^{-1-1/\beta}\,\sigma^{2+\alpha/\beta}$ ($M/L\propto
R_e^{0.22}\,\sigma^{0.49}$; J\o rgensen et al.\ 1996). Despite a large
scatter, this is an important statement, relating the stellar
populations of spheroids to their structural parameters. It is
probably unaffected by dark matter, since spheroids appear baryon
dominated within one $R_e$ (to be contrasted with the Tully-Fisher
relation for spirals; Freeman, these proceedings).

A relation between $\sigma$ and the line-strength index Mg$_2$ (or
broadband colors) also exists, varying slightly among clusters,
possibly due to age or most likely metallicity variations (J\o rgensen
et al.\ 1996). $M/L$ thus varies between clusters (bad for distances
if not accounted for), and bulges and ellipticals do not all have the
same probability distribution of characteristic parameters. The
scatter in the Mg$_2$ ``FP'' is also real, implying some scatter in
the stellar populations.

%
%
\begin{figure}
\plottwo{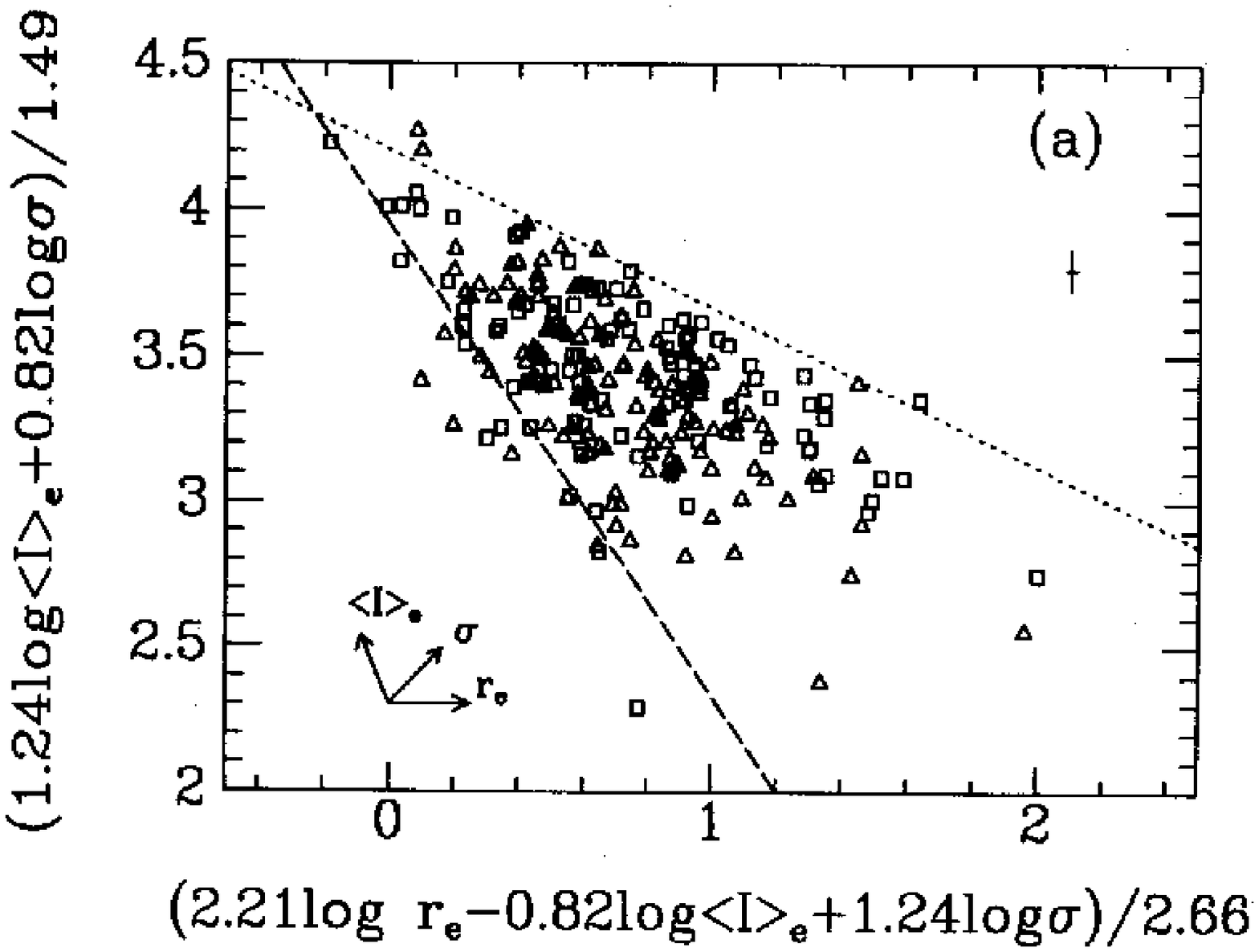}{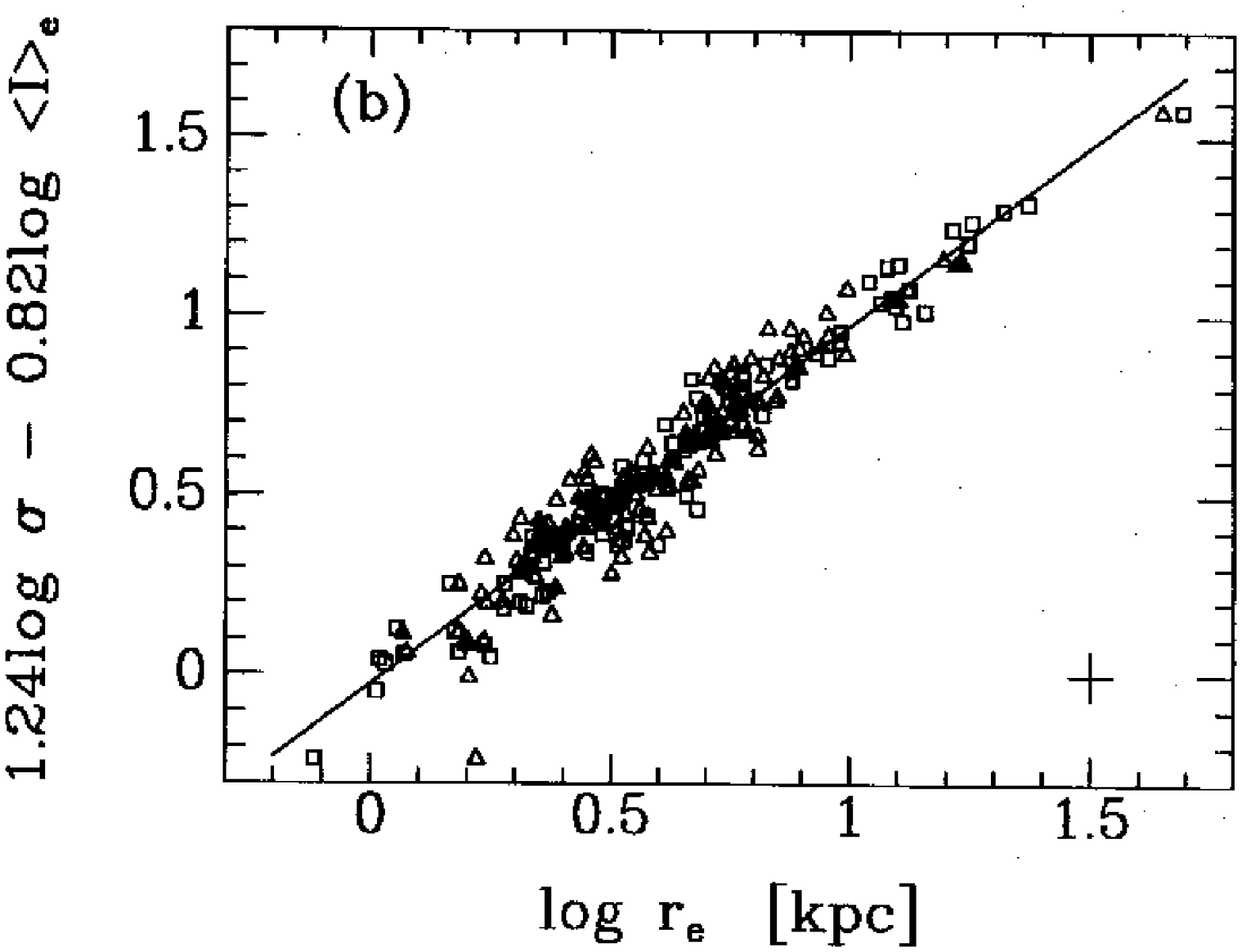}
\caption{Fundamental plane of spheroids in Gunn~$r$. {\em Left:}
Face-on view. The dashed line represents a selection effect (limiting
magnitude), while the dotted line does not. {\em Right:} Edge-on
view. {\em Boxes:} Ellipticals. {\em Triangles:} Bulges (S0s). Typical
error bars are shown in each panel. Reproduced with permission from
J\o rgensen et al.\ (1996).}
\end{figure}

Although ``early'' bulges populate the FP slightly differently from
ellipticals (as do compact ellipticals, dwarf ellipticals, and dwarf
spheroidals), they follow the same Mg$_2$--$\sigma$ relation, and
their properties again suggests a continuation of the elliptical
sequence, indicative to some of a merging sequence with varying
degrees of dissipation (see Bender, Burstein, \& Faber 1993).
%
%
\section{Light Distribution of Bulges\label{sec:light}}
\vspace*{-2mm} So far, we have assumed structural homology. Sersic's
(1968) law allows to characterize structural differences between
galaxies using the the shape parameter $n\,$:
$I(r)\propto\exp[-(r/r_{_0})^{1/n}]$, where $I(r)$ is the surface
brightness profile and $r_{_0}$ some characteristic radius. For $n=4$
and $n=1$, we retrieve the usual $R^{1/4}$ and exponential
profiles. Imposing the $R^{1/4}$ law leads to biased measurements of
$R_e$ and $I_e$ (and $\sigma$) and affects the tilt of the FP; the
departure from virial expectation is reduced when using Sersic
profiles and deprojected quantities.

Galaxies and bulges are better fitted by Sersic law and show a great
variety of shapes: $n$ extends from over 10 to 0.5 as one goes from
brightest cluster galaxies to normal ellipticals and S0s, bulges, and
dwarfs (Graham et al.\ 1996). For bulges alone, $n$ varies
systematically from 6 to 1 from early to late-type systems (high to
low bulge-to-disk ratio), with weaker trends as a function of
luminosity and size (see Fig.~3; Andredakis, Peletier, \& Balcells
1995). This again suggests a similar formation mechanism for all
spheroids, as different mechanisms for early and late-type systems (or
normal and pseudo bulges, see below) would likely lead to a bimodal
distribution of $n$.

In violent relaxation, galaxies with deep central potentials lead to
high $n$ (Hjorth \& Madsen 1995); conversely, the central potential
increases with $n$ (Ciotti 1991; both for spherical isotropic
models). For bulges, the formation or interaction with the disk may
affect the density distribution. The continua of properties mentioned
above thus somewhat support the suggestion that all spheroids harbor a
disk (e.g.\ Burstein et al.\ 2001), although their influence is
probably small in large ellipticals and there are few indications of
disks in dwarfs.

Kormendy (1993) also argues that a number of bulges, referred to as
pseudo-bulges, show structural and kinematic evidence for disk-like
dynamics. These include: i) $\sigma$ smaller than expected from the FJ
relation; ii) fast rotation, with $V/\sigma$ above the isotropic
oblate rotator line in the $V/\sigma-\epsilon$ diagram; iii) bulges as
flat as the disk; iv) spiral structure within $R^{1/4}$ profiles; and
v) substantial population~I material in later types. This suggests
that pseudo-bulges may really be high surface brightness central
disks, and that disks may have a steeper inner light profile than the
inward extrapolation of an exponential. A transition from bulge to
disk-dominated properties is suggested at types Sb--Sbc.

%
%
\begin{figure}
\plotone{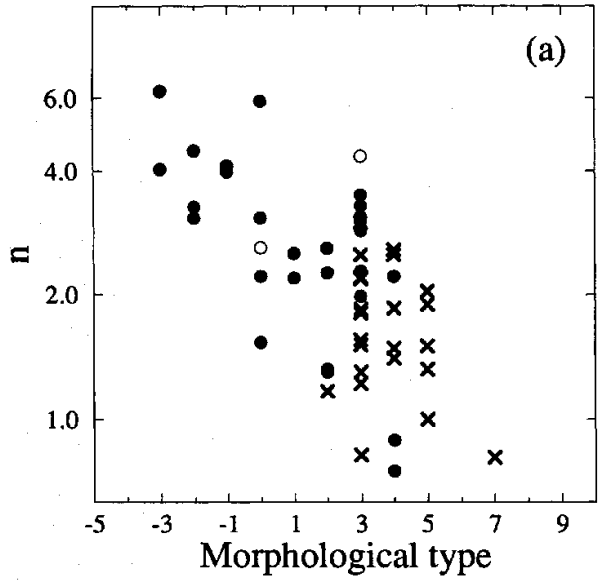}
\caption{Shape of bulges. Sersic's (1968) shape parameter $n$ plotted as a
function of the host galaxy morphological type. {\em Circles:}
Andredakis et al.\ (1995) sample. {\em Crosses:} Kent (1986)
sample. Open symbols represent barred galaxies. No error bars are
plotted for clarity. Reproduced with permission from Andredakis et
al.\ (1995).}
\end{figure}

This picture is consistent with simulations of gas flow in barred
galaxies, which lead to high (and flat) central gas concentrations,
possibly feeding a central BH and forming stars (e.g.\ Friedli
\& Benz 1993, Heller \& Shlosman 1994). The bulge and disk
scalelengths also correlate, independently of type, further suggesting
that at least some bulges grow secularly out of disk material (de Jong
1996). Evolution is then more than the simple aging of the stellar
populations.
%
%
%
\section{Three-Dimensional Structure of Bulges\label{sec:3D}}
\vspace*{-3mm} 
We have argued that bulges are oblate spheroids, but it has long been
known that many bulges show a boxy or peanut-shaped (B/PS) morphology
(e.g.\ Shaw 1987). The vast majority of these are probably bars seen
edge-on. N-body models show that shortly after a bar forms, it
buckles, thickens, and appears almost round, peanut, or boxy when seen
end-on, side-on, or at an intermediate angle. The evolution strongly
depends on the (dark) halo-to-disk mass ratio, but simulations always
result in a B/PS bulge with an exponential vertical light profile
(e.g.\ Combes et al.\ 1990). The thickening is probably due to
vertical heating of the disk through resonant scattering of orbits by
the bar (vertical inner Lindblad resonance (ILR)):
$\Omega_p=\Omega-\nu_z/2$, where $\Omega_p$ is the bar pattern speed
and $\Omega$ and $\nu_z$ the stellar rotation and vertical oscillation
frequencies. The vertical and horizontal ILRs also converge, so that
$\kappa\approx\nu_z$ where the maximum thickening occurs ($\kappa$ is the
epicyclic frequency), and the peanut shape is sustained by orbits
trapped around the 3D generalization of the (2D) $x_1$ family.

B/PS bulges are not found preferentially in groups or clusters, but
they do show an increase of nearby companions. Although accretion
(soft merging) can lead, in principle, to B/PS bulges (Binney \&
Petrou 1985), it probably accounts only for a minor fraction, perhaps
related to the ``thick boxy bulges'' of L\"utticke \& Dettmar
(1999). Hybrid scenarios, where the formation of a bar is triggered by
an interaction, are also possible (Mihos et al.\ 1995). $N$-body
models show cylindrical rotation in the inner parts of B/PS bulges, as
suggested by the few observations available (e.g.\
NGC4565\index{object, NGC4565}; Kormendy \& Illingworth 1982), the
fraction of B/PS bulges and barred disks are consistent (L\"{u}tticke,
Dettmar, \& Pohlen 2000), and B/PS bulges show plateaus in their light
profiles. However, to prove that B/PS bulges are related to bars, and
are thus triaxial, one really wants to probe the potential, requiring
kinematics in the bulge region.

Periodic orbits provide a zeroth order view of stellar
kinematics. Because of the non-homogeneous distribution of the orbits
(see, e.g., Contopoulos \& Gr\o sbol 1989), clear signatures of
non-axisymmetry are seen in the position-velocity diagrams (PVDs) of
edge-on barred disks (Bureau \& Athanassoula 1999). But stars can move
on trapped or chaotic orbits, washing out PVD substructures, and more
realistic $N$-body models indeed reveal subtler signatures (Bureau \&
Athanassoula, in preparation). Gas, however, responds very strongly to
a non-axisymmetric potential. Shocks along the bar cause inflow,
deplete the gas in the outer bar regions, and lead to characteristic
gaps in the PVDs (if a nuclear spiral is formed, requiring an ILR;
Kuijken \& Merrifield 1995, Athanassoula \& Bureau 1999). Line-ratios
can also help identify bars (shock versus photoionization). Merrifield
\& Kuijken (1999) and Bureau \& Freeman (1999) applied these
diagnostics and showed an almost one-to-one correspondence between
B/PS bulges and large-scale bars (Fig.~4), although a few cases may be
due to accretion. The strength of the bar also correlates with the
boxiness of the isophotes. Thus, contrary to ellipticals, where it is
caused by anisotropic velocity dispersions, triaxiality in bulges is
due to high rotation (bar instability).

As face-on galaxies often show photometrically distinct bars and
(rounder) bulges, it is still unclear whether the above thick bars are
truly one with the bulge, or whether a more axisymmetric bulge is
simply buried within them. A complete 3D picture of barred galaxies
(and bulges) is thus still missing.
%
%
\section{Nuclear Properties of Bulges\label{sec:nuclear}}
\vspace*{-3mm} 
While the large-scale structure and kinematics of nearby bulges can be
studied from the ground, HST is required to reach scales of a few tens
of parsecs. An HST/WFPC2 study of a large sample of bulges (mostly
unbarred, Sa--Sbc) by Carollo et al.\ (1997, 1998) reveals a large
variety of nuclear properties, even among early types. Some
``classical'' bulges exist, but in half the cases a bulge is not even
clearly detected. i) Many early-type galaxies show no evidence for a
smooth bulge (also dust lanes, spiral structure, etc); ii) 30\% of
bulges have an irregular central bright component with scattered star
forming regions. Other nuclear star formation occurs, sometimes in
ring-like structures, but it is unclear whether it is associated with
the bulge or inner disk; iii) for types later than S0/a, half the
objects have a resolved, compact central source (often associated with
an elongated structure), the luminosity of which correlates with that
of the host galaxy but not the type (typically brighter in star
forming objects); iv) the brightest compact sources appear similar to
young star clusters in the $M_V-R_e$ plane, while fainter sources are
intermediate between ellipticals and $R^{1/4}$ bulges and globular
clusters, possibly indicating an age sequence. Those sources are
photometrically distinct from their surroundings and are not a simple
steepening of the light profile. These facts suggest a late formation
epoch for some bulges, possibly in disk-driven dissipative accretion
events.

%
%
\begin{figure}
\plotone{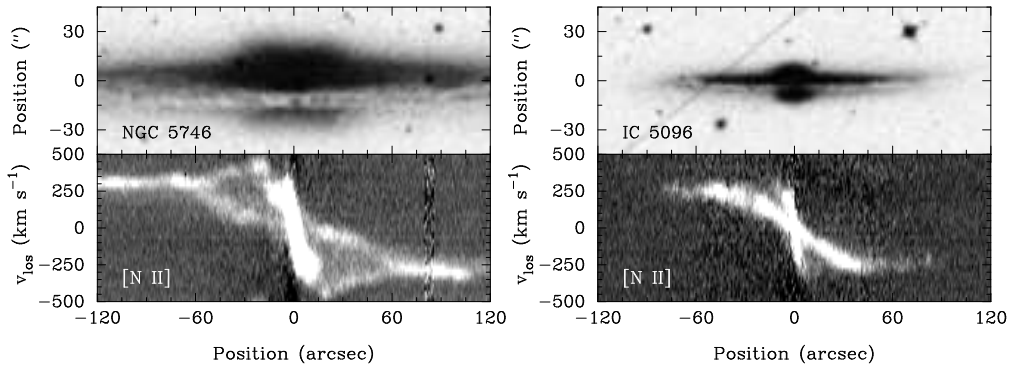}
\caption{Bar signature in B/PS bulges. Image and ionized gas PVD (on
the same scale and along the major-axis) for two B/PS bulges. {\em
Left:} NGC5746\index{object NGC5746}, probably seen side-on. {\em
Right:} IC5096\index{object IC5096}, probably seen end-on. Adapted
from Bureau \& Freeman (1999) with permission.}
\end{figure}

The nuclear light profiles of spheroids (bulges and ellipticals) is
well described by the cusp slope $\gamma$ ($I(r)\propto r^{-\gamma}$
as $r\!\!\rightarrow\!\!0$; Byun et al.\ 1996). In the above sample,
$R^{1/4}$-like bulges have cusps and nuclear densities similar to
ellipticals (at a given spheroid luminosity $L_s$), and also steeper
cusp slopes as $L_s$ is lowered (Faber et al.\ 1997, Carollo \&
Stiavelli 1998). Exponential-like bulges show the same (weaker)
dependence on $L_s$, but they have smaller cusps and nuclear densities
at a given luminosity (Fig.~5). As a group, they thus break the
general trend among spheroids of increasing density with decreasing
luminosity, a rare indication for a different formation
mechanism. This does not indicate a simple evolution along
the Hubble sequence, however, as it holds true for a given type.

%
%
\begin{figure}
\plottwo{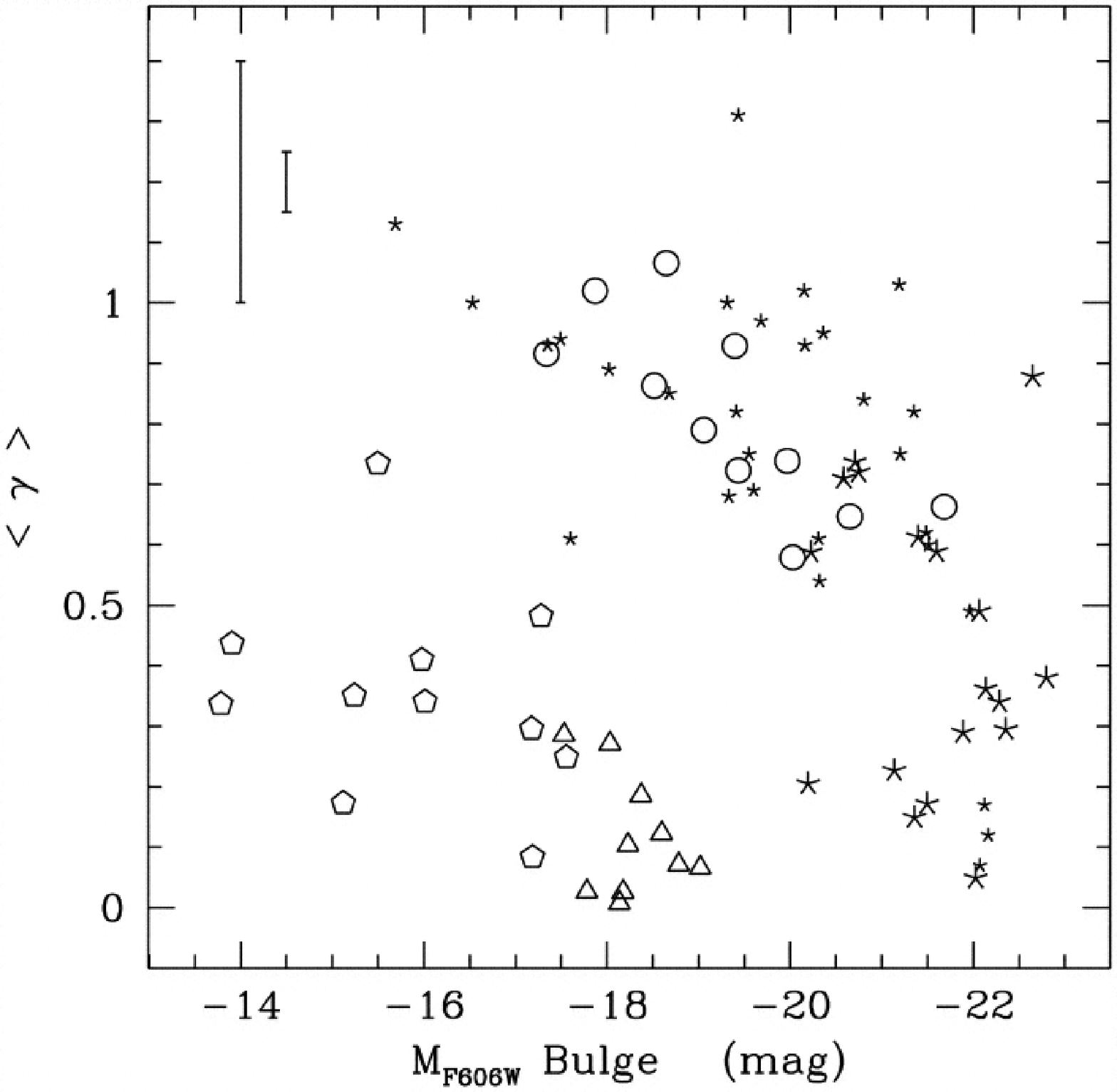}{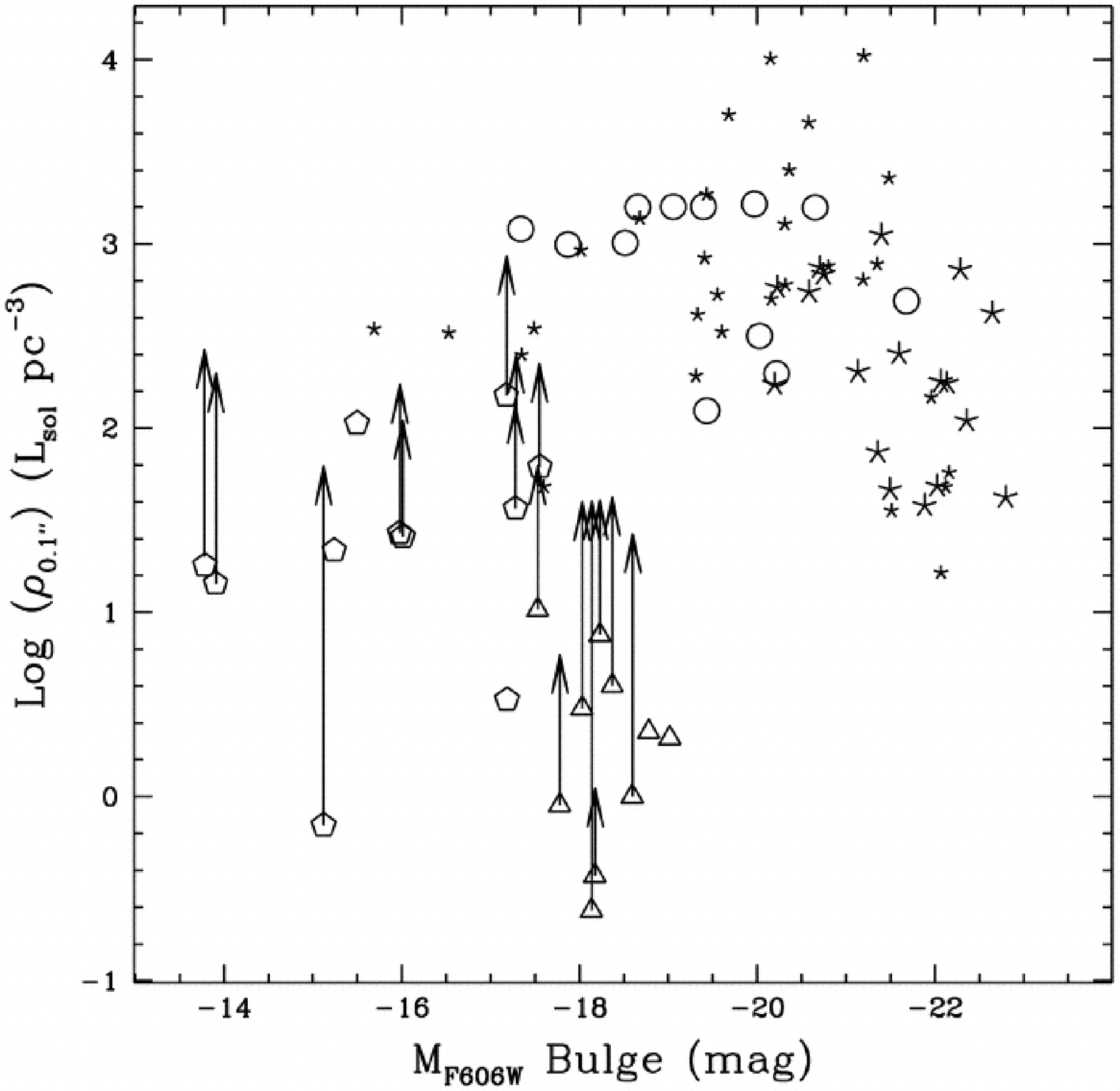}
\caption{Nuclear cusp slopes and densities of spheroids. {\em Left:}
Average logarithmic cusp slope (0\farcs1--0\farcs5) versus
spheroid absolute magnitude. Error bars are shown for galaxies with
and without a central compact source. {\em Right:} Stellar density at
0\farcs1 (from deprojected analytic fits) versus spheroid absolute
magnitude. {\em Triangles and pentagons:} Exponential-like
bulges. {\em Circles:} $R^{1/4}$-like bulges. {\em Asterisks:}
Ellipticals. Reproduced with permission from Carollo \& Stiavelli
(1998).}
\end{figure}

Magorrian et al.\ (1998) proposed the first central BH mass
$M_\bullet$ to spheroid mass $M_s$ (or spheroid luminosity $L_s$)
relation, showing that power-law galaxies (steep cusps) have smaller
$M_\bullet$ and $M/L$ than core galaxies (shallow cusps). But the
masses, based on ground-based kinematics and two-integral axisymmetric
dynamical models, were overestimated. Using HST/STIS kinematics
(resolving the sphere of influence of the BH) and three-integral
models (allowing velocity anisotropy near the center) reduces the
masses by a few. Essentially all galaxies require
$M_\bullet\sim0.001\,M_s$, suggesting a universal baryon fraction
going in the BH. $M_\bullet$ correlates significantly better with
$L_s$ than the total galactic luminosity, indicating that BHs are not
related to disks. The correlation is also independent of bulge type
($R^{1/4}$, exponential, pseudo), suggesting a close link between BH
and bulge (spheroid) formation, independently of how the latter
proceeds.

The relation between $M_\bullet$ and (some measure of) the central
velocity dispersion $\sigma$ is much tighter, although its exact
dependence is debated: $M_\bullet\propto\sigma^{3.8-4.8}$, with a
steep slope favored (Gebhardt et al.\ 2000a, Ferrarese \& Merritt
2000, Merritt \& Ferrarese 2001). The scatter of $\approx0.3$ dex (at
fixed $\sigma$) is consistent with observational errors, indicating
negligible intrinsic scatter. The previous relation with $M_s$ can now
be ``understood'', since $M_s\proptosim L_s^{5/4}$ (Faber et al.\
1987) and $L_s\proptosim\sigma^4$ (FJ), hence $M_s\proptosim\sigma^5$
also. Bulges (spheroids) can now be seen as populating a 2D plane in a
4D space ($\log M_\bullet$, $\log R_e$, $\log\sigma$, $\log L$), the
$M_\bullet-\sigma$ relation being an edge-on projection of this plane
(while the $M_\bullet-L_s$ relation is not, thus the larger
scatter). BH masses predicted from the $M_\bullet-\sigma$ relation are
consistent with reverberation mapping measurements in active galactic
nuclei (AGN; Gebhardt et al.\ 2000b, Ferrarese et al.\ 2001),
indicating a close relationship between quiescent and active BHs, and
strengthening the link between BHs, AGN, and bulge (spheroid)
formation (e.g.\ Silk \& Rees 1998).

There have been many suggestions that bars can be destroyed by central
masses, secularly building bulges over many generations, and moving
galaxies along the Hubble sequence (e.g.\ Norman, Sellwood, \& Hasan
1996). BH masses reported are however an order of magnitude lower than
required ($\sim\!0.1$\% of $M_s$ rather than a few). The central stellar
clusters discussed above do have the right masses, but they would
prevent bars from (re-)forming in late-type exponential bulges,
inhibiting their growth and evolution into early-type $R^{1/4}$ bulges
unless there is a substantial accretion of cold material (the same
applies to BHs). The omnipresence of bars ($\ga70$\% of galaxies;
e.g.\ Seiger \& James 1998) implies a very fast duty cycle, however,
which seems unlikely. At the moment, the evidence is thus against bar
(destruction)-driven secular evolution in bulges.
%
%
\begin{figure}
\plotone{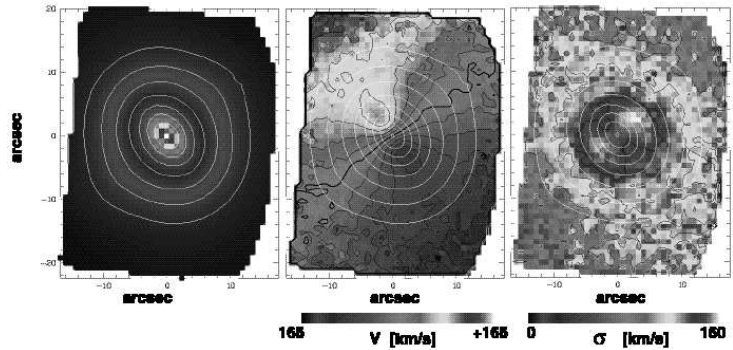}
\caption{SAURON stellar kinematics of the SB0 galaxy
NGC3384\index{object, NGC3384}. {\em Left:} Reconstructed intensity
map. {\em Center:} Velocity. {\em Right:} Velocity dispersion. Not
shown are the Gauss-Hermite moments $h_3$ and $h_4$ (skewness and
kurtosis of the velocity profiles), and the line-strength indices
H$\beta$, Mg$_b$, Fe5015, and Fe5270 (stellar populations). The data
clearly reveal a confined, cold kinematic component inside the
bulge. Reproduced with permission from de Zeeuw et al.\ (2002).}
\end{figure}
%
%
\vspace*{-3mm}
\section{Large-Scale Mass Distribution of Bulges\label{sec:large-scale}}
\vspace*{-2mm} 
There is little to say about the (very) large-scale mass distribution
of bulges, i.e.\ their dark matter content, as little is known. When
there is neutral hydrogen and/or ionized gas, the usual kinematic
tracers in disk galaxies, it often has a complicated geometry or is
disturbed. As shown by Capaccioli et al.\ (1993), it is extremely hard
to push traditional long-slit spectroscopy with integrated light to
significant radii ($\ga2R_e$). They report for NGC3115\index{object,
NGC3115} an increase in $M/L$ from 6 to more than 10 as $r$ goes from
1 to 2$R_e$. Contrary to elliptical galaxies (e.g.\ Hui et al.\ 1995
for Cen~A), there has been little use of globular clusters (GCs) and
planetary nebulae (PN) as (stellar) tracers in bulges. Using GCs in
NGC3115, Kavelaars (1998) shows that $M/L\approx19$ at $5R_e$,
suggesting the presence of dark matter in the halo. This is assumed to
be generic but should be verified. The GC system also shows a red
rapidly rotating metal rich thick disk system and a blue slowly
rotating metal poor halo system, which is not unusual. Results from
polar-rings and other objects probing the potential perpendicular to
the equatorial plane are discussed in detail by Sparke (these
proceedings).
%
%
\section{Conclusions and New Perspectives\label{sec:conclusions}}
\vspace*{-2mm}
Since this is a review, conclusions will be short. Already in the
early 1980s, a continuum between bright ellipticals, low-luminosity
ellipticals, and bright bulges had been demonstrated (Davies et al.\
1983). Now, a continuity in the structural and kinematic properties of
bright bulges (generally early and $R^{1/4}$-like) and faint bulges
(generally late and exponential-like) also emerges (e.g.\ Andredakis
et al.\ 1995). A link between faint bulges and disks is even suggested
and is the subject of much work (Kormendy 1993, these
proceedings). All the observations discussed in this paper concern
nearby galaxies, where the internal structure, kinematics, and
dynamics can be studied in detail. This shows that so-called
near-field cosmology has an essential (and perhaps dominant) role to
play in our quest to understand galaxy formation and evolution.

Although it is impossible to be exhaustive, it is essential to discuss
current instrumental developments, since they will lead without doubt
to the next discoveries. The usual bells and whistles associated with
``weather'' prediction are however necessary. On nuclear scales, we are
unlikely to make great advances from space until the advent of NGST
(e.g.\ Stockman \& Mather 2000). HST/ACS does not increase HST's
spatial resolution, and HST/STIS will not be upgraded or replaced,
offering few new possibilities for high spatial resolution
kinematics. Adaptive optics on large ground-based telescope,
particularly in the near-infrared and/or with integral-field
spectrographs (IFSs), is very promising (e.g.\ VLT/SINFONI; Mengel et
al.\ 2000). On intermediate scales, WHT/SAURON has already
demonstrated the possibilities of wide-field IFSs, especially when
supplemented with data on nuclear scales and proper modeling tools
(Fig.~6; de Zeeuw et al.\ 2002). VLT/VIMOS and other similar
instruments will increase this power. On large scales, WHT/PNS will
provide much needed data on stellar kinematics in the outskirts of
galaxies (using PN as tracers; see Douglas \& Taylor 1999),
constraining the amount of dark matter present. Astrometric missions
such as ESA/GAIA (e.g.\ Perryman et al.\ 2001), acting on all scales,
will provide the position, colors, type, and radial velocity for a
billion stars in the Galaxy (a large fraction with accurate proper
motions and parallax). This will revolutionize our view of spirals,
providing us with a stereoscopic and kinematic census of the stellar
populations. Near-field cosmology at its best!
%
%
\acknowledgements Support for this work was provided by NASA through
Hubble Fellowship grant HST-HF-01136.01 awarded by the Space Telescope
Science Institute, which is operated by the Association of
Universities for Research in Astronomy, Inc., for NASA, under contract
NAS~5-26555. All figures reproduced by permission of the American
Astronomical Society and Blackwell Publishing.
%
%

%
\end{document}